\documentclass[a4paper]{article}
\usepackage{ASVspoof}
\usepackage{epsfig,amssymb,amsmath}
\usepackage{hyperref}
\usepackage{multirow}
\usepackage{makecell}
\usepackage{multicol, subcaption, balance} 
\usepackage{mathtools, nccmath}
\usepackage{booktabs}
\usepackage{enumitem}
\usepackage{color}
\usepackage{adjustbox}
\usepackage{comment}
\usepackage{bigdelim} 
\ninept

\title{BUT Systems and Analyses for the ASVspoof 5 Challenge}

\makeatletter
\def\name#1{\gdef\@name{#1\\}}
\makeatother
\name{{\em  Johan Rohdin$\textsuperscript{1,2}$, Lin Zhang$\textsuperscript{1}$, Oldřich Plchot$\textsuperscript{1}$, Vojtěch Staněk$\textsuperscript{1}$, David Mihola$\textsuperscript{1}$, Junyi Peng$\textsuperscript{1}$,}\\
      {\em Themos Stafylakis\textsuperscript{2}, Dmitriy Beveraki$\textsuperscript{2}$, Anna Silnova$\textsuperscript{1}$, Jan Brukner$\textsuperscript{1}$, Lukáš Burget$\textsuperscript{1}$}}

\address{$\textsuperscript{1}$Brno University of Technology, $\textsuperscript{2}$Omilia Conversational Intelligence \\
{\small \tt \{rohdin, qzhang, iplchot\}@fit.vutbr.cz},
{\small \tt tstafylakis@omilia.com}
}
\begin{document}

\newcommand{\zlin}[1]{\textcolor{blue}{#1}}

\maketitle

\begin{abstract}
This paper describes the BUT submitted systems for the ASVspoof 5 challenge, along with analyses. For the conventional deepfake detection task, we use ResNet18 and self-supervised models for the closed and open conditions, respectively. In addition, we analyze and visualize different combinations of speaker information and spoofing information as label schemes for training. For spoofing-robust automatic speaker verification (SASV), we introduce effective priors and propose using logistic regression to jointly train affine transformations of the countermeasure scores and the automatic speaker verification scores in such a way that the SASV LLR is optimized.
\end{abstract}

\section{Introduction}
Automatic speaker verification (ASV) systems are widely used to verify the identity of speakers. However, ASV systems are also vulnerable to spoofing attacks~\cite{evans2013spoofing, Ring20219,10.1145/3477314.3507013}. Although the main purpose of generative models is to facilitate people's lives, not to attack biometric models or present false information, the advancement of generative models poses an increased threat to biometric systems and society. Thus, it is desirable to explore anti-spoofing systems (also known as countermeasures - CM or presentation attack detection - PAD) to detect and prevent spoofing attacks. To encourage researchers to work on this important task in the speech processing field, the ASVspoof \cite{Wu2015_asvspoof2015, Kinnunen2017_asvspoof2017, Nautsch2021_asvspoof2019, junichi2022_asvspoof2021} challenge has been held since 2015. So far, three different spoofing scenarios have been discussed in previous years: (1) physical access (PA) for replay attacks, (2) logical access (LA) for spoofed speech generated by text-to-speech (TTS) synthesis and/or voice conversion (VC) attacks, and (3) deepfake (DF) for strongly compressed LA attacks.

This year's ASVspoof 5 \cite{Wang2024_ASVspoof5} involves two tracks. Track 1, like in previous years, involves conventional deepfake detection that discriminates bona fide speech from spoofed speech. For the main discussed spoofing scenario, ASVspoof 5 combined LA and DF based on advanced TTS/VC systems and introduced adversarial attacks (specifically focusing on Malafide \cite{panariello23b_Malafide} and its upgraded version Malacopula \cite{Max2024_Malacopula}) to the unseen evaluation set. Track 2 in ASVspoof 5 merged with the spoofing-robust automatic speaker verification (SASV) \cite{jung22c_sasv2022} 2022 challenge, with a newly defined task-agnostic metric, a-DCF \cite{shim24_aDCF}. Conveniently, newly proposed score fusion based on a non-linear combination of CM and ASV log-likelihood ratios (LLRs) after score calibration \cite{wangRevisiting2024} and a single integrated system based on SKA-TDNN \cite{mun23_ska-tdnn} are provided as baselines. The challenge also defines two conditions in each track - close and open - depending on whether it is allowed to use external data.

For track 1 close condition, we followed top-ranked teams from previous years and utilized ResNet18 as our submitted system. Furthermore, we explored the influence of training with speaker labels in combination with spoofed/bona fide labels. As for the open condition, given the promising performance of SSL models for spoof detection \cite{wang22_odyssey, tak22_odyssey, Kawa2023}, we compared different SSL models as front-end. We utilized our previously proposed Multi-head Factorized Attentive Pooling (MHFA) \cite{peng2022} to efficiently aggregate information from transformer layers through an attention mechanism, which showed superior results compared to a simple pooling layer.

For track 2, there are two common approaches: (1) score/embedding fusion \cite{shim22_sasv22_fusion, wang22ea_interspeech} between two independent ASV and CM, or (2) a single integrating model \cite{zeng22_interspeech, alenin22_sasv2_IDRD} that optimizes ASV and CM simultaneously. Among these, score fusion is more widely used as it is more intuitive and makes the model decision more explainable. 
Most existing score fusion studies focus on simple score summation.
However, in order to take a decision that minimizes the expected cost for a trial, a non-linear combination of the CM and ASV is necessary~\cite{wangRevisiting2024}. In this work, we provide a more general treatment of the SASV scoring problem. We derive the general SASV LLR and show that the optimal SASV decision for any choice of cost parameters can be taken from the SASV LLR where the priors have been replaced by \emph{effective priors}, which are obtained by absorbing the costs of various incorrect decisions into the original priors. We then use these results to jointly optimize the calibration of the CM and ASV scores to provide an accurate SASV LLR.

\section{System for track 1 deepfake detection}\label{sec:track1}
\subsection{Task and database}
Track 1 of ASVspoof 5 focuses on the stand-alone speech deepfake detection task, distinguishing bona fide samples from spoofed ones, just like in previous competitions \cite{Wu2015_asvspoof2015, Kinnunen2017_asvspoof2017, Nautsch2021_asvspoof2019, junichi2022_asvspoof2021}.

The data of this challenge were collected during ASVspoof 5 Phase 1. The overall dataset is based on the Multilingual Librispeech (MLS) dataset (English-language subset) \cite{pratap20_MLS}, and synthetic data are collected from community volunteers. 
Eight, eight, and sixteen spoofing attacks are considered for training, development, and evaluation sets, respectively. There are 400 and 785 speakers involved in the training and development set respectively. To measure the performance of deepfake detection, the minimum detection cost function (minDCF), the cost of log-likelihood ratio ($C_\text{llr}$) \cite{BRUMMER2006230_cllr}, and the equal error rate (EER) are considered in track 1 of the challenge. More details can be found in the summary paper of the challenge \cite{Wang2024_ASVspoof5}.

\subsection{ResNet18 for the closed condition}
\subsubsection{Details of the system}\label{sec:deepfake_close:detail}
For the closed condition, we chose ResNet18 \cite{he2016_resnet} as our system with MUSAN \cite{snyder2015musan} noise subset and room impulse responses for data augmentation, as it is the most used system by top-ranked teams in the previous years \cite{liu2023asvspoof2021summary}. We use 80-dimensional Mel-filterbank with a window length of 25 ms and a frame shift of 10 ms. After extracting embedding from ResNet18, we use the temporal statistics pooling layer, a linear layer with 256 units, a ReLU activation function, a batch normalization layer, and another linear layer with softmax activation for calculating cross-entropy loss with K-class classification. The Likelihoods for bonafide/spoof were computed by summing the likelihoods over speakers and spoof types where applicable, after which the LLR were computed. This approach may not be ideal but due to time constraints we did not explore alternative approaches. In the next subsection, we will analyze different K based on whether and how we utilize speaker information for classification.

\subsubsection{Comparison on different labeling schemes}

To explore whether speaker information could help deepfake detection in the ASVspoof 5 challenge, we analyzed different label schemes considering different speaker identity and bona fide/spoof classes in this subsection. This is motivated by conflicting conclusions in existing studies.
Some studies propose that simultaneously optimizing speaker classification and deepfake detection would enhance the robustness of deepfake detection \cite{Mo2022_mtl_3auxiliaryloss}. Whereas others claim that reducing speaker variability 
would be beneficial for deepfake detection \cite{Gajan2020_adversial}. We examined five types of labeling schemes based on ResNet18 introduced in the previous subsection. Three (\texttt{spk-binspf}, \texttt{spk-mulspf}, \texttt{spk-onespf}) of these schemes include speaker identity, while the other two (\texttt{mulspf}, \texttt{binspf}) focus on bona fide/spoof(s) classification, as shown below. The numbers of classes K for each are annotated, given there are four hundred speakers, eight different spoofing methods (A01 $\sim$ A08), and one bona fide in the training set.

\begin{itemize}[itemsep=0mm, leftmargin=4mm, nosep]
    \item \texttt{spk-binspf} (K = 800): The label is a combination of speaker ID and bona fide/spoof,
    \item \texttt{spk-mulspf} (K = 3600): The label is a combination of speaker ID and bona fide/A01/.../A08,
    \item \texttt{spk-onespf} (K = 401): The label is speaker ID in case of bona fide, else ``spoof,''  
    \item \texttt{mulspf} (K = 9): The label is bona fide/A01/.../A08,
    \item \texttt{binspf} (K = 2): The label is bona fide/spoof, the same as in common deepfake detection.
\end{itemize}

\begin{figure*}[!t]
\centering
\includegraphics[width=1.84\columnwidth]{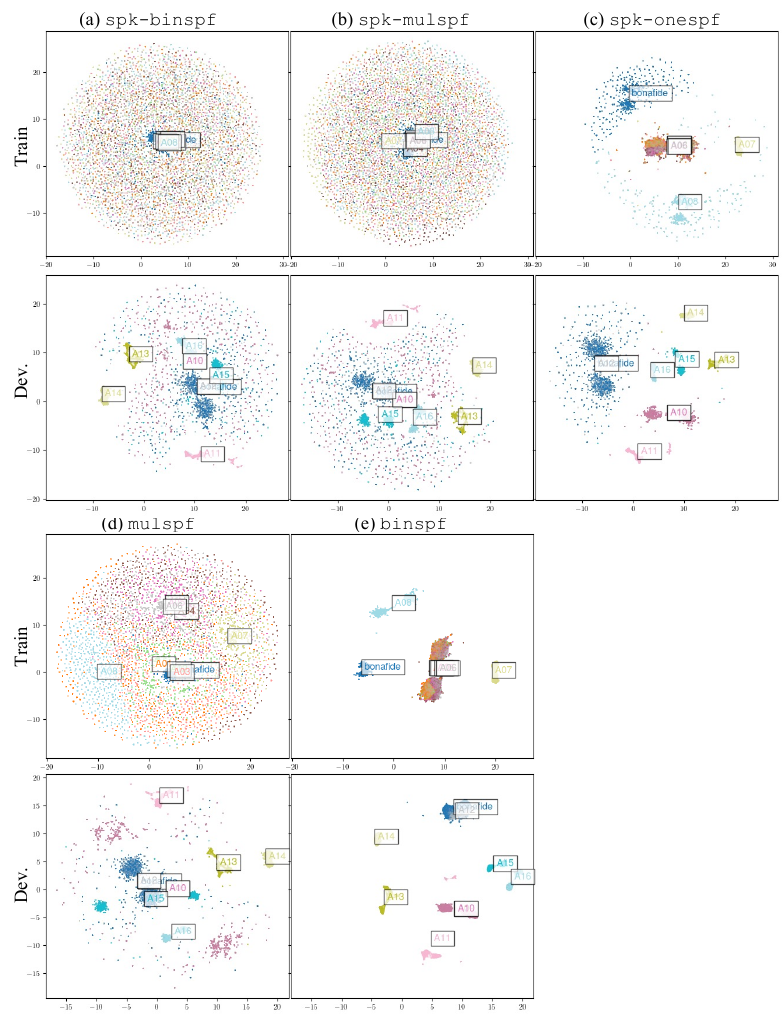}
\caption{{\it Embedding space of different label schemes.}} \label{fig:emb_space}
\vspace{-2mm}
\end{figure*}

Results using the above five label schemes on the development set of track 1 are shown in Table \ref{tab:res_5label_track1}, and visualization of their embedding spaces by UMAP \cite{mcinnes2018umap-software} are shown in Figure \ref{fig:emb_space}. 

First, we focus on the training set regarding the seen scenario to discuss three questions:

\noindent\textit{(1) Should we consider speaker information for spoofed speech?} Figure \ref{fig:emb_space} (a) to (c) show the models considering speaker ID. In the (a) \texttt{spk-binspf} and (b) \texttt{spk-mulspf}, which treat spoofing methods for each speaker as independent classes, we observe that although some bona fide samples are clustered in the center, most samples are mixed and difficult to distinguish. In (c) \texttt{spk-onespf}, after we integrate all spoofed samples as a single spoof class without considering ``spoofed'' speaker information, the spoofed speech begins to distinguish from bona fide speech.
This model achieves the lowest EER and promising minDCF compared with other models.
This hints that it is challenging to train the model
when assigning speaker ID to the spoofed speech. This could be due to the reduced number of samples for each class or the confusion introduced by speaker information for spoofed speech.

\noindent\textit{(2) Should we consider speaker information for bona fide speech?}
Visualization on comparing Figure \ref{fig:emb_space} (b) vs. (d), and (c) vs. (e) shows that when we remove speaker information and focus on deepfake detection as in (d) and (e), the samples are more clustered according to their bona fide/spoof labels.
This is understandable, as the models in (b) and (c) contain speaker information, and the differences between speaker characteristics might be larger than the differences between spoofed and bona fide samples, which makes it difficult to learn how to distinguish bona fide from spoof. 

\noindent\textit{(3) Should we integrate different spoofing methods?}
Comparing (d) and (e), we can observe that integrating different spoofing methods as a single spoof class helps the model distinguish spoofed samples from bona fide. Given that we didn't treat different spoofing methods independently, it is understandable that A01 to A06 are mixed. Notably, A07 and A08 are still well distinguished from others even though they are trained under the same label. Similar observations can be found in (c-d). This is acceptable, as A07 is generated by FastPitch \cite{lancucki2021fastpitch} and A08 is generated by VITS \cite{kim2021vits}, which are different compared to the other six spoofing methods based on GlowTTS and GradTTS.

Next, we move to the development set, which contains unseen spoofing methods and disjoint speakers. Across all five label schemes in Figure \ref{fig:emb_space} (a-e), we observe that A11 (Tacotron 2 \cite{shen2018tacotron2}), A13 (StarGANv2-VC \cite{li21e_StarGANv2-VC}) and A14 (YourTTS \cite{casanova2022yourtts}) are well distinguished in all label schemes compared with other spoofing methods. This could be because A11, A13, and A14 utilized similar or the same components with spoofing methods that the model encountered during training. For example, Tacotron2 technology applied in A11 is also utilized by A07 (FastPitch \cite{lancucki2021fastpitch}) to estimate the duration of the input symbols. YourTTS used in A14 is built based upon VITS \cite{kim2021vits} that is applied by A08 of the training set. This shows that the model is still limited to the seen scenarios, and its performance is restricted by the degree of mismatch from the training set. 
Meanwhile, A12 (In-house unit-selection) \cite{Wang2024_ASVspoof5} is difficult to distinguish from the bona fide, which is understandable as unit-selection selecting segments from bona fide utterances to consist the desired spoofed speech, and it is unseen during training. More robust technology for detecting such unit-selection attacks is worth exploring in future work. Additionally, exploring adaptation methods to handle mismatched scenarios more effectively is essential for future research.

\begin{table}[t]
\caption{\label{tab:res_track1_resnet18} {\it Results of ResNet18 with different label schemes on the development set of track 1.}}\label{tab:res_5label_track1}
\vspace{-2mm}
\setlength{\tabcolsep}{3pt}
\centering{
\begin{tabular}{cccccc}
\toprule
ID & Model & minDCF & EER (\%) & $C_\text{llr}$    & actDCF  \\
\midrule
1 & \texttt{spk-binspf} & 0.2401  & 13.640      & 2.6596 & 0.9349 \\
2 & \texttt{spk-mulspf} & 0.2708  & 13.818      & 0.7985 & 0.6129 \\
3 &\texttt{spk-onespf} & 0.1624  & 11.891      & 3.1482 & 0.7794 \\
4 & \texttt{mulspf}     & 0.1811  & 12.456      & 5.4365 & 0.9978 \\
5 & \texttt{binspf}     & 0.1374  & 12.156      & 2.6360 & 0.6720 \\
\bottomrule
\end{tabular}}
\vspace{-3mm}
\end{table}

Finally, we submit the system 5 (ResNet18-\texttt{binspf}) that uses bona fide/spoof classes as our final system for the track 1 - close condition. It achieves minDCF=0.5809, actDCF=0.8537, $C_\text{llr}$= 4.0994, and EER=23.34\% in the evaluation set.  

\subsection{Pretrained SSL with MHFA for the open condition}

SSL models have attracted attention in the deepfake speech detection area due to their promising performance ~\cite{wang22_odyssey, tak22_odyssey, Kawa2023}. Therefore, we used pretrained SSL models as our CM system submitted to track 1-open.

Specifically, we compared different pretrained SSL models including Wav2vec2 model \cite{Baevski2020}, WavLM model \cite{Chen2022}, Hubert \cite{hsu2021hubert}, and data2vec \cite{baevski2022data2vec} as shown in Table \ref{tab:res_ssl_track1}. All SSL models are in their Base version given the prohibition using of LibriLight in ASVspoof challenge \cite{Wang2024_ASVspoof5}. In addition, when we aggregated hidden features extracted from transformer layers of SSL models, we compared learnable weighted sum \cite{yang21c_interspeech_superb} with average pooling (AP) vs. MHFA~\cite{peng2022}. MHFA is our recently proposed pooling method that utilizes two sets of normalized layer-wise weights to generate attention maps and compressed features. We followed the configuration from our previous paper \cite{peng2022} with the number of heads set as 64. During training, only the parameters of MHFA while keeping SSL models frozen. No data augmentation was applied in these experiments. Results are shown in Table \ref{tab:res_ssl_track1}.

\begin{table}[t]
\centering
\caption{\it Performances of pretrained SSL models (Base versions) on the development set of Track 1-open condition.}\label{tab:res_ssl_track1}
\vspace{-2mm}
\setlength{\tabcolsep}{2.5pt}
\scalebox{.95}{
\begin{tabular}{cllcccc}
\toprule
ID & \multicolumn{2}{c}{Model}     & minDCF & EER (\%) & $C_\text{llr}$    & actDCF  \\
\midrule
1 & Wav2vec2 &+ AP    & 0.1312  & \phantom{0}5.094    & 0.2924 & 0.1674 \\
2 & Wav2vec2 &+ MHFA  & 0.0848  & \phantom{0}3.300    & 0.5876 & 0.1097 \\
3 & HuBERT &+ MHFA    & 0.2497  & 11.164   & 0.8306 & 1.0000 \\
4 & WavLM &+ MHFA     & 0.1400  & \phantom{0}6.881    & 0.8268 & 1.0000 \\
5 & Data2vec &+ MHFA  & 0.2231  & \phantom{0}8.400    & 0.8067 & 0.2724 \\
\midrule
6 & \multicolumn{2}{l}{Fusion 2 + 4} &  0.0763 & \phantom{0}2.974 & 0.87861 & 1.0000 \\
7 & \multicolumn{2}{l}{Fusion 6 + ResNet18} &  0.0693 & \phantom{0}3.048 & 0.90159 & 1.0000 \\
\bottomrule
\end{tabular}}
\end{table}

Comparing systems 1 and 2, MHFA outperformed simple AP. Comparing 2 $\sim$ 5, Wav2vec2 and WavLM achieve better performance. Thus, we submitted the system 7 -- equal weight averaging of max-min normalized prediction scores from systems 2, 4 from Table \ref{tab:res_ssl_track1} and system 5 from Table \ref{tab:res_track1_resnet18}. The fused system 7 achieves minDCF=0.2573, actDCF=1.0000, $C_\text{llr}$=0.9955, EER=9.28\% in the evaluation set.

\section{System for track 2}\label{sec:track2}
\subsection{Task and database}
Track 2 of ASVspoof 5 involves a spoofing-robust automatic speaker verification (SASV) task \cite{jung22c_sasv2022}. This is a newly introduced track that has emerged in recent years, aiming to integrate ASV and CM systems and accepting the speech only if it is spoken by the bona fide target speaker. The training and development sets are the same as in track 1 with the additional well-known Voxceleb2 \cite{chung18b_VoxCeleb2} database from the speaker verification field. Voxceleb2 provided 5,994 speakers for training.
For measuring performance, different from the SASV 2022 \cite{jung22c_sasv2022}, which utilizes SASV-EER as the main metric, this year's challenge uses the newly introduced min a-DCF \cite{shim24_aDCF} as the primary metric, with min t-DCF \cite{kinnunen2020tDCF_TASLP} and t-EER \cite{kinnunen2023tEER} as the supplement metrics. More details can be found in the summary paper of the challenge \cite{Wang2024_ASVspoof5}.

\subsection{ASV systems}
We based our ASV systems on the ResNet architecture by following the exact recipes with the Voxceleb dataset from the Wespeaker toolkit\footnote{\url{https://github.com/wenet-e2e/wespeaker}} ~\cite{wang2023wespeaker} while omitting speech and music parts from the MUSAN data that were not allowed to be used in the challenge. We experimented on ResNet34, ResNet101, and ResNet221 with Additive Angular Margin (AAM) softmax loss. We compared models w/ and w/o large margin fine-tuning. As enrollment embeddings, we used the average embedding from multiple enrollment utterances. We also analyzed normalizing extracted embeddings by subtracting the mean of the ASVspoof5 train set/voxceleb2 dev set. For scoring, we used simple cosine similarity. The results are shown in Table \ref{tab:res_ASV}. 

\label{sec:resnet_asv}

\begin{table}[t]
\caption{\label{tab:res_ASV} {\it EERs (\%) of ASV systems on the development set of Track 2.}}
\vspace{-2mm}
\setlength{\tabcolsep}{3pt}
\centering{
\begin{tabular}{ccccc}
\toprule
ID & Model & Large Margin & Mean Norm. & EER (\%) \\
\midrule
1 & ResNet34\phantom{0}       & \checkmark   & -                   & 5.935        \\
2 & ResNet34\phantom{0}       & \checkmark   & ASVspoof5 train      & 5.605        \\
3 & ResNet34\phantom{0}       & \checkmark   & Voxceleb2 dev.       & 5.952        \\
4 & ResNet34\phantom{0}       & -            & ASVspoof5 train      & 5.464        \\
5 & ResNet101      & -            & ASVspoof5 train      & 5.233        \\
6 & ResNet221      & -            & ASVspoof5 train      & 5.101       \\
\bottomrule
\end{tabular}}
\vspace{-3mm}
\end{table}

\subsection{Spoofing-robust automatic speaker verification system}

Our SASV system is based on combining the CM LLR and the ASV LLR into a SASV LLR, i.e., the LLR for the hypotheses 
\begin{itemize}
\setlength\itemsep{0em}
\item
$\mathcal{H}_A$ (Accept hypothesis): The speech is bona fide and from the target speaker.

\item
$\mathcal{H}_R$ (Reject hypothesis): $\mathcal{H}_A$ is not true. 
\end{itemize}
Since for binary decision problems, the optimal decision can be taken from the LLR, this is the optimal score for a SASV system.  Under some assumptions, the SASV LLR can be computed from the LLR of the CM system, the LLR of the ASV system, and the priors of the cost parameters. The formula for the resulting SASV LLR has been provided before, e.g., in \cite{wangRevisiting2024} and \cite{todisco18_interspeech}. Here, we present a slightly more general form of the LLR and introduce the concept of \emph{effective priors} for the SASV task, inspired by this concept in speaker verification \cite{brümmer2013bosaristoolkittheoryalgorithms}. The use of effective priors shows how to make optimal decisions in the scenarios where the different types of false accept have different costs, which is not the case in this evaluation\footnote{Both false acceptance of an impostor (ASV false accept) and false acceptance of spoofed speech (CM false accept) have cost 10 in this evaluation.} but may be useful in other scenarios and, more importantly for this evaluation, enables us to do fusion/calibration using logistic regression.

The details of our approach are provided in the following subsections. We denote the speech $X$ and the properties of the speech as follows:
\begin{itemize}[itemsep=0mm, leftmargin=4mm, nosep]
\setlength\itemsep{0em}
\item
$B$: speech is bona fide 
\item
$S$: speech is spoofed
\item
$T$: speech is from the target speaker
\item
$N$: speech is not from the target speaker
\end{itemize}
where $B$ and $S$ are disjoint and $T$ and $N$ are disjoint. Note that $\mathcal{H}_A=(B,T)$ and $\mathcal{H}_R$ is the union of $(B,N)$, $(S,T)$ and $(S,N)$\footnote{When we refer to spoofed speech from a target/non-target speaker, we mean spoofed speech with simulated characteristics similar to those of the true target/non-target speaker.}.

\subsubsection{Optimal scores and decisions}
\label{sec:opt_dec}
Given some speech data, $X$, the expected cost of rejecting a trial is $C_{\rm{miss}}P(\mathcal{H}_A|X)$, where $C_{\rm{miss}}$ is the cost of false rejection. If the cost of incorrectly accepting a spoofed utterance, $C_{\rm{fa},spoof}$, and incorrectly accepting an impostor, $C_{\rm{fa,imp}}$\footnote{Note that in the ASVspoof 5 evaluation plan, this is denoted $C_{\rm{fa}}$.}, is the same (as is the case in ASVspoof 5), the
expected cost of accepting a trial is $C_{\rm{fa}}P(\mathcal{H}_R|X)$, where $C_{\rm{fa}}=C_{\rm{fa},spoof}=C_{\rm{fa,imp}}$. The problem is then a standard binary decision theory problem for which the optimal decision is to accept if (see e.g. the BOSARIS toolkit manual~\cite{brümmer2013bosaristoolkittheoryalgorithms})
\setlength{\jot}{8pt}
\begin{align}
\frac{C_{\rm{miss}}P(\mathcal{H}_A|X)}{C_{\rm{fa}}P(\mathcal{H}_R|X)}
=
\frac{C_{\rm{miss}}P(X|\mathcal{H}_A) P(\mathcal{H}_A)  }{C_{\rm{fa}}P(X|\mathcal{H}_R) P(\mathcal{H}_R)} >1 \nonumber\\
\Leftrightarrow
\log\frac{P(X|\mathcal{H}_A)}{P(X|\mathcal{H}_R)}>\log\frac{C_{\rm{fa}}}{C_{\rm{miss}}}-\log \frac{P(\mathcal{H}_A)}{P(\mathcal{H}_R)}.
\end{align} 
Although the evaluation plan did not explicitly ask participants to provide LLRs for track 2, the above shows that using LLRs and an appropriate threshold leads to optimal decisions.\footnote{Strictly speaking, we do not need to provide the threshold, and the LLRs could be subjected to any monotonically rising function and still be optimal for the challenge metric since it does not care about calibration.}

\subsubsection{LLR}
The LLR is given by
\setlength{\jot}{8pt}
\begin{align}
\label{eq:sasv_llr1}
&\log\frac{P(X|\mathcal{H}_A)}{P(X|\mathcal{H}_R)} \nonumber\\
&=\log\frac{P(X|B,T)}{P(X|B,N)p_\text{BN}+ P(X|S,T)p_\text{ST} +  P(X|S,N)p_\text{SN}} \nonumber\\
&=-\log\frac{P(X|B,N)p_\text{BN}+ P(X|S,T)p_\text{ST} + P(X|S,N)p_\text{SN}  } {P(X|B,T)}\nonumber\\
&=-\log\left(\frac{P(X|B,N)}{P(X|B,T)}p_\text{BN} + \frac{P(X|S,T)}{P(X|B,T)}p_\text{ST} \nonumber\right.\\
&\quad\quad\quad\quad\quad\quad\quad\left.+ \frac{P(X|S,N)}{P(X|B,T)}p_\text{SN}\right)
,\end{align}

\noindent where
\begin{itemize}
\setlength\itemsep{0em}
\item
$p_\text{BN} = P(B,N|\mathcal{H}_R)$
\item
$p_\text{ST} = P(S,T|\mathcal{H}_R)$
\item
$p_\text{SN} = P(S,N|\mathcal{H}_R)$
\end{itemize}
while $p_\text{BT} = P(B,T|\mathcal{H}_A)=1$ is kept implicit. The first term after the final equal sign in Eq.~(\ref{eq:sasv_llr1}) is just the inverted LLR for ASV on \emph{bona fide} speech and the second term is just the inverted \emph{speaker-dependent} LLR for a CM. Our CM systems are speaker-independent, which is incorrect according to the above formula but hopefully a good enough approximation.  
Note that the LLR depends on conditional priors, $p_\text{BN}$, $p_\text{ST}$ and $p_\text{SN}$. This is common to LLRs that are composed of several LLRs for simpler \emph{subevents}, see \cite{brümmerCompoundLLR} and \cite{solewicz22_odyssey} for two other examples. In many SASV scenarios (including the challenge if we understand correctly), we do not expect spoofing of non-target speakers, i.e., $p_\text{SN}=0$.

\subsubsection{Effective priors}
Analogously to common practice in ASV~\cite{brümmer2013bosaristoolkittheoryalgorithms}, it simplifies matters to convert the cost parameters into an \emph{equivalent} set of costs parameters\footnote{Contrary to Section \ref{sec:opt_dec}, we here also include spoofed impostors. The cost of false acceptance of such trials is denoted $C_{\rm{fa,spoof,imp}}$.} where $C_{\rm{miss}}=C_{\rm{fa},spoof}=C_{\rm{fa,imp}}=C_{\rm{fa,spoof,imp}}=1$ but where the priors are replaced with the \emph{effective priors}. With general costs, we shall accept the trial if 
\setlength{\jot}{10pt}%
\begin{align}
\label{eq:effective_priors1}
\frac{
P(X|B,T)P(B,T)
C_{\rm{miss}}
}
{
\splitfrac{
P(X|B,N)P(B,N)
C_{\rm{fa,imp}}
}
{
\splitfrac{
+ 
P(X|S,T)P(S,T)
C_{\rm{fa,spoof}}
}
{
+ 
P(X|S,N)P(S,N)
C_{\rm{fa,spoof,imp}}
}
}
}
>1.
\end{align}
By dividing the numerator and denominator by  
\setlength{\jot}{2pt}%
\begin{align}
Z&= P(B,T) C_{\rm{miss}}  
 + P(B,N)C_{\rm{fa,imp}} \nonumber\\
 &+ P(S,T)C_{\rm{fa,spoof}} 
 + P(B,N)C_{\rm{fa,spoof,imp}},
 \end{align}
we obtain
\setlength{\jot}{10pt}%
\begin{align}
\label{eq:effective_priors2}
\frac{
P(X|B,T)P'(B,T)
}
{
\splitfrac{
P(X|B,N)P'(B,N)
}
{
\splitfrac{
+ 
P(X|S,T)P'(S,T)
}
{
+ 
P(X|S,N)P'(S,N)
}
}
}
>1,
\end{align}
where, e.g.,
\begin{equation}
P'(B,N)=C_{\rm{fa,imp}} P(B,N)/Z,
\end{equation}
and the other \emph{effective priors}, $P'(B,T)$, $P'(S,T)$ and $P'(S,N)$ are defined similarly.
This means that the decision that is optimal according to Ineq.~(\ref{eq:effective_priors2}) is also optimal according to Eq.~(\ref{eq:effective_priors1}) and vice versa. Thus, we can work with $C_{\rm{miss}}=C_{\rm{fa},spoof}=C_{\rm{fa,imp}}=C_{\rm{fa,spoof,imp}}=1$ and the original priors replaced by the effective priors, $P'(\cdot,\cdot)$, when taking decisions as well as when optimizing models for taking decisions such as calibration and fusion models. In this way, we can handle situations when originally, e.g., $C_{\rm{fa},spoof}\ne C_{\rm{fa,imp}}$. In the case when $C_{\rm{fa},spoof}= C_{\rm{fa,imp}}$, working with effective priors is still helpful for training, e.g., calibration models (see the next subsubsection).

\subsubsection{Logistic regression based calibration/fusion}
Let $p_\text{SN}=0$ and denote 
\setlength{\jot}{1pt}%
\begin{align}
{\rm{llr}}_{\rm{cm}}(X)=\log\frac{P(X|B,T)}{P(X|S,T)}
\end{align}
and
\begin{align}
{\rm{llr}}_{\rm{asv}}(X)=\log\frac{P(X|B,T)}{P(X|B,N)},
\end{align}
then 
\setlength{\jot}{2pt}%
\begin{align}
\label{eq:sasv_llr3}
\rm{llr}&_{\rm{sasv}}(X) =\log\frac{P(X|\mathcal{H}_A)}{P(X|\mathcal{H}_R)}\nonumber\\
&=-\log\left(p'_\text{BN}e^{-{\rm{llr}}_{\rm{cm}}(X)} + p'_\text{ST}e^{-{\rm{llr}}_{\rm{asv}}(X)} \right)
\end{align}
We note that although the primary metric, min a-DCF, is calibration insensitive, i.e., it does not care whether the SASV LLR is calibrated, proper calibration of the CM and ASV LLR is still important for min a-DCF due to the complex relation between them and the SASV LLR given by Eq.~(\ref{eq:sasv_llr3}). Calibrating the raw CM and ASV LLRs (denoted $\rm{llr}_{\rm{asv}}^{\rm{raw}}$ and $\rm{llr}_{\rm{cm}}^{\rm{raw}}$) with affine transformations, we obtain the \emph{corrected} SASV LLR  
\setlength{\jot}{1pt}%
\begin{align}
\label{eq:sasv_llr4}
&{\rm{llr}}_{\rm{sasv}}(X,\tilde{a}_0,\tilde{a}_1,\tilde{c}_0,\tilde{c}_1)\nonumber\\
&=-\log\left(p'_\text{BN}e^{-\tilde{c}_1{\rm{llr}}_{\rm{cm}}^{\rm{raw}}(X)-\tilde{c}_0} + p'_\text{ST}e^{-\tilde{a}_1{\rm{llr}}_{\rm{asv}}^{\rm{raw}}(X)-\tilde{a}_0} \right).
\end{align}
We then learn the calibration parameters $\tilde{a}_0$, $\tilde{a}_1$, $\tilde{c}_0$ and $\tilde{c}_1$ jointly with logistic regression with the three classes, $(B,T)$, $(B,N)$ and $(S,T)$, being weighted according to their effective priors, i.e.,  
\setlength{\jot}{1pt}%
\begin{align}
&\tilde{a}_0,\tilde{a}_1,\tilde{c}_0,\tilde{c}_1=\nonumber\\
 &\arg\min_{\substack{a_0,a_1,\\c_0,c_1}}
 \sum_{D}\frac{P'(D)}{N_{D}} \sum_{i=1}^{N_{D}}L(X,S_D,a_0,a_1,c_0,c_1)
\end{align}
where
\begin{align}
\label{eq:sasv_llr_optim}
L&(X,S_D,a_0,a_1,c_0,c_1)
\nonumber\\
&=\log\left(1+e^{-S_D\left(
 {\rm{llr}}_{\rm{sasv}}(X,a_0,a_1,c_0,c_1) + \tau\right)}\right) 
 ,
\end{align}
$D=\{ (B,T),(B,N),(S,T)\}$, $N_D$ is the number of trials for class $D$,
\begin{equation}
S_D =\left\{ 
\begin{aligned}
& \phantom{x}1 &&  {\rm{for}}\, (B,T) \\
& -1 && {\rm{for}}\,(B,N)\,{\rm{and}}\,(S,T) \\
\end{aligned}
\right.
\end{equation}
and
\begin{equation}
\tau=\log\frac{P'(B,T)}{P'(B,N)+P'(S,T)}.
\end{equation}
Logistic regression on LLR is a standard approach for calibration and fusion in speaker verification \cite{4291590} which encourages good calibration as well as discrimination.

\subsubsection{Experiments}
We implemented the logistic regression calibration in Pytorch~\cite{NEURIPS2019_9015}. For optimization, we used its L-BFGS~\cite{Liu_1989} optimizer with default settings. The result using \texttt{spk-binspf} of Track one as CM LLR and system 2 of Table~\ref{tab:res_track1_resnet18} the ASV LLR are presented in Table~\ref{tab:track2}. We can see that the proposed calibration improves the min a-DCF with around 1\% absolute. 
\begin{table}[t]
\caption{\label{tab:track2} {\it Impact of calibration for the SASV LLR of track 2 in terms of min a-DCF on the ASVspoof5 development set. \emph{No calibration} refers to combining the CM and ASV LLR according Eq.~(\ref{eq:sasv_llr3}), i.e., without any calibration and \emph{Calibration} refers to combining the CM and ASV LLR according Eq.~(\ref{eq:sasv_llr4}) with the calibration parameters optimized according to Eq.~(\ref{eq:sasv_llr_optim}).}}
\vspace{-2mm}
\setlength{\tabcolsep}{3pt}
\centering{
\begin{tabular}{cc}
\toprule
No calibration & Calibration \\
\midrule
0.17874  & 0.16854      \\
\bottomrule
\end{tabular}}
\end{table}
Due to time constraints, we have not evaluated the effect of calibration for systems other than  \texttt{spk-binspf}, nor have we compared it with alternative approaches for combining the CM and ASV LLR. A brief discussion of some of the conceptual aspects is provided in the next subsection while further experimental evaluations and analysis should be part of future work.

The results for the closed condition of track 2 are shown in Table~\ref{tab:res_closed_track2}.
\begin{table}[t!]
\centering
\caption{\it Results of various systems on the closed condition of track 2 on the ASVspoof5 development set. The number under the \emph{CM system} heading refers to the ID in Table~\ref{tab:res_track1_resnet18} and the number under the \emph{ASV system} heading refers to the ID in Table~\ref{tab:res_ASV}.}\label{tab:res_closed_track2}
\setlength{\tabcolsep}{2.5pt}
\vspace{-2mm}
\scalebox{.95}{
\begin{tabular}{clccc}
\toprule
CM system & ASV system  & min a-DCF & min t-DCF & t-EER (\%)  \\
\midrule
1 & 2  & 0.16854 & 0.35234 & 8.196 \\
5 & 2  & 0.12696 & 0.20858 & 6.569 \\
5 & 5  & 0.12529 & 0.20858 & 5.977 \\
5 & 6  & 0.12527 & 0.20924 & 6.026 \\
\bottomrule
\end{tabular}}
\end{table}
Consistent with the results for track 1, we can see that CM system 5 outperforms CM system 1. The ASV system has a very minor impact on min a-DCF but a larger impact on t-EER. For the closed condition of track 2, we submitted the system in the last row. It achieves min a-DCF=0.389, min t-DCF=0.778, t-EER=20.850\% in the evaluation set.

The results for the open condition of track 2 are shown in Table~\ref{tab:res_open_track2}.
\begin{table}[t!]
\centering
\caption{\it Results of various systems on the open condition of Track 2 on the ASVspoof5 development set. The number under the \emph{CM system} heading refers to the ID in Table~\ref{tab:res_ssl_track1} and the number under the \emph{ASV system} heading refers to the ID in Table~\ref{tab:res_ASV}.}\label{tab:res_open_track2}
\vspace{-2mm}
\setlength{\tabcolsep}{2.5pt}
\scalebox{.95}{
\begin{tabular}{clccc}
\toprule
CM system & ASV system  & min a-DCF & min t-DCF & t-EER (\%)  \\
\midrule
2 & 6 & 0.04761 & 0.18886 & 2.514 \\
7 & 6 & 0.07287 & 0.15573 & 2.026 \\
\bottomrule
\end{tabular}}
\end{table}
Contrary to track 1, the fusion of several CM systems did not improve in the primary metric for track 2. Due to limited time, we did not explore this further and submitted the system in the first row.  It achieves min a-DCF=0.180, min t-DCF=0.543, t-EER=8.390\% in the evaluation set.

\subsubsection{Discussion}
Equation~ (\ref{eq:sasv_llr3}) is the same as 
\cite{wangRevisiting2024} and \cite{todisco18_interspeech}. In those papers, it was suggested to tune $p'_\text{BN}$ and $p'_\text{ST}$ with a grid search. In addition, discriminative calibration of the CM and ASV LLR was done individually before combining them to form the SASV LLR in \cite{wangRevisiting2024}. In \cite{todisco18_interspeech}, both the ASV LLR and the CM LLR were estimated by a generative (Gaussian) fusion of the raw CM and raw ASV scores. However we keep $p'_\text{BN}$ and $p'_\text{ST}$ as specified by the cost parameters (the effective versions) and instead jointly learn affine transformations of $\rm{llr}_{\rm{cm}}$ and $\rm{llr}_{\rm{asv}}$ that optimizes the SASV LLR on the left side of Eq.~(\ref{eq:sasv_llr3}). 
A few points can be made:
\begin{itemize}[itemsep=0mm, leftmargin=4mm, nosep]
\setlength\itemsep{0em}
\item
Most calibration methods rarely produce scores that are well-calibrated at all operating points. Joint optimization of the CM and ASV calibration should calibrate these LLRs to be optimal for the operating point of the SASV task. This speaks in favor of our proposed method.
\item
Tuning $p'_\text{BN}$ and $p'_\text{ST}$ corresponds to adjusting the offsets of the CM and ASV LLR. This is less powerful than affine transformations. However, individual precalibration of the CM and ASV LLR followed by tuning of $p'_\text{BN}$ and $p'_\text{ST}$  for the SASV as in \cite{wangRevisiting2024} and \cite{todisco18_interspeech} task could be sufficient.
\item
Tuning parameters with a grid search as in \cite{wangRevisiting2024} and \cite{todisco18_interspeech} allows optimizing the performance of DCF at one specific operating point. Since the DCF of one operating point is not a continuous function, it cannot be optimized by gradient-based methods such as L-BFGS.
\item
Logistic regression corresponds to optimizing the calibration for a wide range of operating points instead of the one specified by DCF \cite{BRUMMER2006230_cllr}. This could make the score less optimized for the specific operating point but, on the other hand, reduce the risk of overfitting to this specific operating point.
\end{itemize}
The pros and cons of different calibration/fusion methods need to be analyzed in future work.

\section{Conclusion}\label{sec:conclusion}
This paper described BUT systems and analyses for the ASVspoof 5 challenge. For track 1, we constructed ResNet18 with analyses on different speaker and spoofing label schemes for the close condition and pretrained SSL model with MHFA for the open condition. For track 2, we defined SASV LLR in a more general form with an introduced concept of effective priors. Introducing effective priors enables optimal decision for the SASV task regardless of cost parameters. It also enables calibrating SASV LLRs as well as evaluating the quality of such LLRs with calibration sensitive metrics.

    \section{Acknowledgements}
This work was partly supported by the European Union's Horizon Europe grant agreement No. 101135916 ``ELOQUENCE,'' and by the Czech Ministry of Interior project No. VB02000060 ``NABOSO.''
We acknowledge VSB – Technical University of Ostrava, IT4Innovations National Supercomputing Center, Czech Republic, for awarding this project access to the LUMI supercomputer, owned by the EuroHPC Joint Undertaking, hosted by CSC (Finland) and the LUMI consortium through the Ministry of Education, Youth and Sports of the Czech Republic through the e-INFRA CZ (grant ID: 90254).

\bibliographystyle{IEEEbib}
\balance
\bibliography{main}

\begin{thebibliography}{10}

\bibitem{evans2013spoofing}
Nicholas Evans, Tomi Kinnunen, and Junichi Yamagishi,
\newblock ``Spoofing and countermeasures for automatic speaker verification.,''
\newblock in {\em Proc. Interspeech}, 2013, pp. 925--929.

\bibitem{Ring20219}
Tim Ring,
\newblock ``Europol: the ai hacker threat to biometrics,''
\newblock {\em Biometric Technology Today}, vol. 2021, no. 2, pp. 9 – 11, 2021.

\bibitem{10.1145/3477314.3507013}
Anton Firc and Kamil Malinka,
\newblock ``The dawn of a text-dependent society: deepfakes as a threat to speech verification systems,''
\newblock in {\em Proc. ACM/SIGAPP SAC}, 2022, p. 1646–1655.

\bibitem{Wu2015_asvspoof2015}
Zhizheng Wu, Tomi Kinnunen, Nicholas Evans, Junichi Yamagishi, Cemal Hanil{\c{c}}i, Md~Sahidullah, and Aleksandr Sizov,
\newblock ``{ASVspoof 2015: the First Automatic Speaker Verification Spoofing and Countermeasures Challenge},''
\newblock in {\em Proc. Interspeech}, 2015, pp. 2037--2041.

\bibitem{Kinnunen2017_asvspoof2017}
Tomi Kinnunen, Md~Sahidullah, H{\'{e}}ctor Delgado, Massimiliano Todisco, Nicholas Evans, Junichi Yamagishi, and Kong~Aik Lee,
\newblock ``{The ASVspoof 2017 Challenge: Assessing the Limits of Replay Spoofing Attack Detection},''
\newblock in {\em Proc. Interspeech}, 2017, pp. 2--6.

\bibitem{Nautsch2021_asvspoof2019}
Andreas Nautsch, Xin Wang, Nicholas Evans, Tomi~H. Kinnunen, Ville Vestman, Massimiliano Todisco, Héctor Delgado, Md~Sahidullah, Junichi Yamagishi, and Kong~Aik Lee,
\newblock ``{ASV}spoof 2019: Spoofing countermeasures for the detection of synthesized, converted and replayed speech,''
\newblock {\em IEEE Transactions on Biometrics, Behavior, and Identity Science}, vol. 3, no. 2, pp. 252--265, 2021.

\bibitem{junichi2022_asvspoof2021}
Junichi Yamagishi, Xin Wang, Massimiliano Todisco, Md~Sahidullah, Jose Patino, Andreas Nautsch, Xuechen Liu, Kong~Aik Lee, Tomi Kinnunen, Nicholas Evans, and Héctor Delgado,
\newblock ``{ASVspoof 2021: accelerating progress in spoofed and deepfake speech detection},''
\newblock in {\em Proc. ASVspoof Workshop}, 2021, pp. 47--54.

\bibitem{Wang2024_ASVspoof5}
Xin Wang, H{\'e}ctor Delgado, Hemlata Tak, Jee-weon Jung, Hye-jin Shim, Massimiliano Todisco, Ivan Kukanov, Xuechen Liu, Md~Sahidullah, Tomi Kinnunen, Nicholas Evans, Kong~Aik Lee, and Junichi Yamagishi,
\newblock ``{ASVspoof 5}: {Crowdsourced} speech data, deepfakes, and adversarial attacks at scale,''
\newblock in {\em Proc. ASVspoof Workshop 2024 (accepted)}.

\bibitem{panariello23b_Malafide}
Michele Panariello, Wanying Ge, Hemlata Tak, Massimiliano Todisco, and Nicholas Evans,
\newblock ``{Malafide: a novel adversarial convolutive noise attack against deepfake and spoofing detection systems},''
\newblock in {\em Proc. Interspeech}, 2023, pp. 2868--2872.

\bibitem{Max2024_Malacopula}
Massimiliano Todisco, Michele Panariello, Xin Wang, Hector Delgado, Kong-Aik Lee, and Nicholas Evans,
\newblock ``{Malacopula: Adversarial automatic speaker verification attacks using a neural-based generalised hammerstein model},''
\newblock in {\em Proc. ASVspoof Workshop 2024 (accepted)}.

\bibitem{jung22c_sasv2022}
Jee weon Jung, Hemlata Tak, Hye jin Shim, Hee-Soo Heo, Bong-Jin Lee, Soo-Whan Chung, Ha-Jin Yu, Nicholas Evans, and Tomi Kinnunen,
\newblock ``{SASV 2022: The First Spoofing-Aware Speaker Verification Challenge},''
\newblock in {\em Proc. Interspeech}, 2022, pp. 2893--2897.

\bibitem{shim24_aDCF}
Hye jin Shim, Jee weon Jung, Tomi Kinnunen, Nicholas Evans, Jean-François Bonastre, and Itshak Lapidot,
\newblock ``{a-DCF: an architecture agnostic metric with application to spoofing-robust speaker verification},''
\newblock in {\em Proc. Odyssey}, 2024, pp. 158--164.

\bibitem{wangRevisiting2024}
Xin Wang, Tomi Kinnunen, Lee Kong~Aik, Paul-Gauthier No{\'e}, and Junichi Yamagishi,
\newblock ``Revisiting and improving scoring fusion for spoofing-aware speaker verification using compositional data analysis,''
\newblock in {\em Proc. {{Interspeech}}}, 2024, p. (accepted).

\bibitem{mun23_ska-tdnn}
Sung~Hwan Mun, Hye jin Shim, Hemlata Tak, Xin Wang, Xuechen Liu, Md~Sahidullah, Myeonghun Jeong, Min~Hyun Han, Massimiliano Todisco, Kong~Aik Lee, Junichi Yamagishi, Nicholas Evans, Tomi Kinnunen, Nam~Soo Kim, and Jee weon Jung,
\newblock ``{Towards Single Integrated Spoofing-aware Speaker Verification Embeddings},''
\newblock in {\em Proc. Interspeech}, 2023, pp. 3989--3993.

\bibitem{wang22_odyssey}
Xin Wang and Junichi Yamagishi,
\newblock ``{Investigating Self-Supervised Front Ends for Speech Spoofing Countermeasures},''
\newblock in {\em Proc. Odyssey}, 2022, pp. 100--106.

\bibitem{tak22_odyssey}
Hemlata Tak, Massimiliano Todisco, Xin Wang, Jee weon Jung, Junichi Yamagishi, and Nicholas Evans,
\newblock ``{Automatic Speaker Verification Spoofing and Deepfake Detection Using Wav2vec 2.0 and Data Augmentation},''
\newblock in {\em Proc. Odyssey}, 2022, pp. 112--119.

\bibitem{Kawa2023}
Piotr Kawa, Marcin Plata, Michał Czuba, Piotr Szymański, and Piotr Syga,
\newblock ``{Improved DeepFake Detection Using Whisper Features},''
\newblock in {\em Proc. Interspeech}, 2023, pp. 4009--4013.

\bibitem{peng2022}
Junyi Peng, Oldřich Plchot, Themos Stafylakis, Ladislav Mošner, Lukáš Burget, and Jan Černocký,
\newblock ``An attention-based backend allowing efficient fine-tuning of transformer models for speaker verification,''
\newblock in {\em Proc. SLT}, 2023, pp. 555--562.

\bibitem{shim22_sasv22_fusion}
Hye jin Shim, Hemlata Tak, Xuechen Liu, et~al.,
\newblock ``{Baseline Systems for the First Spoofing-Aware Speaker Verification Challenge: Score and Embedding Fusion},''
\newblock in {\em Proc. Odyssey}, 2022, pp. 330--337.

\bibitem{wang22ea_interspeech}
Xingming Wang, Xiaoyi Qin, Yikang Wang, Yunfei Xu, and Ming Li,
\newblock ``{The DKU-OPPO System for the 2022 Spoofing-Aware Speaker Verification Challenge},''
\newblock in {\em Proc. Interspeech}, 2022, pp. 4396--4400.

\bibitem{zeng22_interspeech}
Chang Zeng, Lin Zhang, Meng Liu, and Junichi Yamagishi,
\newblock ``{Spoofing-Aware Attention based ASV Back-end with Multiple Enrollment Utterances and a Sampling Strategy for the SASV Challenge 2022},''
\newblock in {\em Proc. Interspeech}, 2022, pp. 2883--2887.

\bibitem{alenin22_sasv2_IDRD}
Alexander Alenin, Nikita Torgashov, Anton Okhotnikov, Rostislav Makarov, and Ivan Yakovlev,
\newblock ``{A Subnetwork Approach for Spoofing Aware Speaker Verification},''
\newblock in {\em Proc. Interspeech}, 2022, pp. 2888--2892.

\bibitem{pratap20_MLS}
Vineel Pratap, Qiantong Xu, Anuroop Sriram, Gabriel Synnaeve, and Ronan Collobert,
\newblock ``{MLS: A Large-Scale Multilingual Dataset for Speech Research},''
\newblock in {\em Proc. Interspeech}, 2020, pp. 2757--2761.

\bibitem{BRUMMER2006230_cllr}
Niko Brümmer and Johan {du Preez},
\newblock ``Application-independent evaluation of speaker detection,''
\newblock {\em Computer Speech \& Language}, vol. 20, no. 2, pp. 230--275, 2006.

\bibitem{he2016_resnet}
Kaiming He, Xiangyu Zhang, Shaoqing Ren, and Jian Sun,
\newblock ``Deep residual learning for image recognition,''
\newblock in {\em Proc. CVPR}, 2016, pp. 770--778.

\bibitem{snyder2015musan}
David Snyder, Guoguo Chen, and Daniel Povey,
\newblock ``{MUSAN}: A music, speech, and noise corpus,''
\newblock {\em arXiv preprint arXiv:1510.08484}, 2015.

\bibitem{liu2023asvspoof2021summary}
Xuechen Liu, Xin Wang, Md~Sahidullah, Jose Patino, H{\'e}ctor Delgado, Tomi Kinnunen, Massimiliano Todisco, Junichi Yamagishi, Nicholas Evans, Andreas Nautsch, et~al.,
\newblock ``{ASV}spoof 2021: Towards spoofed and deepfake speech detection in the wild,''
\newblock {\em IEEE/ACM Transactions on Audio, Speech, and Language Processing}, vol. 31, pp. 2507--2522, 2023.

\bibitem{Mo2022_mtl_3auxiliaryloss}
Yichuan Mo and Shilin Wang,
\newblock ``Multi-task learning improves synthetic speech detection,''
\newblock in {\em Proc. ICASSP}, 2022, pp. 6392--6396.

\bibitem{Gajan2020_adversial}
Gajan Suthokumar, Vidhyasaharan Sethu, Kaavya Sriskandaraja, and Eliathamby Ambikairajah,
\newblock ``Adversarial multi-task learning for speaker normalization in replay detection,''
\newblock in {\em Proc. ICASSP}, 2020, pp. 6609--6613.

\bibitem{mcinnes2018umap-software}
Leland McInnes, John Healy, Nathaniel Saul, and Lukas Grossberger,
\newblock ``{UMAP: U}niform manifold approximation and projection,''
\newblock {\em The Journal of Open Source Software}, vol. 3, no. 29, pp. 861, 2018.

\bibitem{lancucki2021fastpitch}
Adrian {\L}a{\'n}cucki,
\newblock ``Fastpitch: Parallel text-to-speech with pitch prediction,''
\newblock in {\em Proc. ICASSP}, 2021, pp. 6588--6592.

\bibitem{kim2021vits}
Jaehyeon Kim, Jungil Kong, and Juhee Son,
\newblock ``Conditional variational autoencoder with adversarial learning for end-to-end text-to-speech,''
\newblock in {\em Proc. ICML}, 2021, pp. 5530--5540.

\bibitem{shen2018tacotron2}
Jonathan Shen, Ruoming Pang, Ron~J Weiss, Mike Schuster, Navdeep Jaitly, Zongheng Yang, Zhifeng Chen, Yu~Zhang, Yuxuan Wang, Rj~Skerrv-Ryan, et~al.,
\newblock ``Natural tts synthesis by conditioning wavenet on mel spectrogram predictions,''
\newblock in {\em Proc. ICASSP}, 2018, pp. 4779--4783.

\bibitem{li21e_StarGANv2-VC}
Yinghao~Aaron Li, Ali Zare, and Nima Mesgarani,
\newblock ``{StarGANv2-VC: A Diverse, Unsupervised, Non-Parallel Framework for Natural-Sounding Voice Conversion},''
\newblock in {\em Proc. Interspeech}, 2021, pp. 1349--1353.

\bibitem{casanova2022yourtts}
Edresson Casanova, Julian Weber, Christopher~D Shulby, Arnaldo~Candido Junior, Eren G{\"o}lge, and Moacir~A Ponti,
\newblock ``Your{TTS}: Towards zero-shot multi-speaker tts and zero-shot voice conversion for everyone,''
\newblock in {\em Proc. ICML}, 2022, pp. 2709--2720.

\bibitem{Baevski2020}
Alexei Baevski, Henry Zhou, Abdelrahman Mohamed, and Michael Auli,
\newblock ``wav2vec 2.0: a framework for self-supervised learning of speech representations,''
\newblock in {\em Proc. NeurIPS}, 2020, p.~12.

\bibitem{Chen2022}
Sanyuan Chen, Chengyi Wang, Zhengyang Chen, and et. al.,
\newblock ``Wav{LM: L}arge-scale self-supervised pre-training for full stack speech processing,''
\newblock {\em IEEE Journal of Selected Topics in Signal Processing}, vol. 16, no. 6, pp. 1505–1518, Oct. 2022.

\bibitem{hsu2021hubert}
Wei-Ning Hsu, Benjamin Bolte, Yao-Hung~Hubert Tsai, Kushal Lakhotia, Ruslan Salakhutdinov, and Abdelrahman Mohamed,
\newblock ``Hubert: Self-supervised speech representation learning by masked prediction of hidden units,''
\newblock {\em IEEE/ACM transactions on audio, speech, and language processing}, vol. 29, pp. 3451--3460, 2021.

\bibitem{baevski2022data2vec}
Alexei Baevski, Wei-Ning Hsu, Qiantong Xu, Arun Babu, Jiatao Gu, and Michael Auli,
\newblock ``Data2vec: A general framework for self-supervised learning in speech, vision and language,''
\newblock in {\em Proc. ICML}, 2022, pp. 1298--1312.

\bibitem{yang21c_interspeech_superb}
Shu wen Yang, Po-Han Chi, and et. al.,
\newblock ``{SUPERB: Speech Processing Universal PERformance Benchmark},''
\newblock in {\em Proc. Interspeech}, 2021, pp. 1194--1198.

\bibitem{chung18b_VoxCeleb2}
Joon~Son Chung, Arsha Nagrani, and Andrew Zisserman,
\newblock ``{VoxCeleb2: Deep Speaker Recognition},''
\newblock in {\em Proc. Interspeech}, 2018, pp. 1086--1090.

\bibitem{kinnunen2020tDCF_TASLP}
Tomi Kinnunen, H{\'e}ctor Delgado, Nicholas Evans, Kong~Aik Lee, Ville Vestman, Andreas Nautsch, Massimiliano Todisco, Xin Wang, Md~Sahidullah, Junichi Yamagishi, et~al.,
\newblock ``Tandem assessment of spoofing countermeasures and automatic speaker verification: Fundamentals,''
\newblock {\em IEEE/ACM Transactions on Audio, Speech, and Language Processing}, vol. 28, pp. 2195--2210, 2020.

\bibitem{kinnunen2023tEER}
Tomi~H Kinnunen, Kong~Aik Lee, Hemlata Tak, Nicholas Evans, and Andreas Nautsch,
\newblock ``{t-EER: Parameter-free tandem evaluation of countermeasures and biometric comparators},''
\newblock {\em IEEE Transactions on Pattern Analysis and Machine Intelligence}, 2023.

\bibitem{wang2023wespeaker}
Hongji Wang, Chengdong Liang, Shuai Wang, Zhengyang Chen, Binbin Zhang, Xu~Xiang, Yanlei Deng, and Yanmin Qian,
\newblock ``Wespeaker: A research and production oriented speaker embedding learning toolkit,''
\newblock in {\em Proc. ICASSP}. IEEE, 2023, pp. 1--5.

\bibitem{todisco18_interspeech}
Massimiliano Todisco, Héctor Delgado, Kong~Aik Lee, Md~Sahidullah, Nicholas Evans, Tomi Kinnunen, and Junichi Yamagishi,
\newblock ``{Integrated Presentation Attack Detection and Automatic Speaker Verification: Common Features and Gaussian Back-end Fusion},''
\newblock in {\em Proc. Interspeech}, 2018, pp. 77--81.

\bibitem{brümmer2013bosaristoolkittheoryalgorithms}
Niko Brümmer and Edward de~Villiers,
\newblock ``The {BOSARIS T}oolkit: Theory, algorithms and code for surviving the new {DCF},''
\newblock {\em arXiv preprint arXiv:1304.2865}, 2013.

\bibitem{brümmerCompoundLLR}
Niko Brümmer,
\newblock ``{LLR transformation for SRE'12},''
\newblock {\em Agnitio Research}, 2012.

\bibitem{solewicz22_odyssey}
Yosef Solewicz, Noa Cohen, Johan Rohdin, Srikanth Madikeri, and Jan~Honza Čercnocký,
\newblock ``{Speaker Recognition on Mono-Channel Telephony Recordings},''
\newblock in {\em Proc. Odyssey}, 2022, pp. 193--199.

\bibitem{4291590}
Niko Brummer, Lukas Burget, Jan Cernocky, Ondrej Glembek, Frantisek Grezl, Martin Karafiat, David~A. van Leeuwen, Pavel Matejka, Petr Schwarz, and Albert Strasheim,
\newblock ``Fusion of heterogeneous speaker recognition systems in the {STBU} submission for the {NIST} speaker recognition evaluation 2006,''
\newblock {\em IEEE Transactions on Audio, Speech, and Language Processing}, vol. 15, no. 7, pp. 2072--2084, 2007.

\bibitem{NEURIPS2019_9015}
Adam Paszke, Sam Gross, et~al.,
\newblock ``Pytorch: An imperative style, high-performance deep learning library,''
\newblock in {\em Proc. NeurIPS}, 2019, pp. 8024--8035.

\bibitem{Liu_1989}
Dong~C Liu and Jorge Nocedal,
\newblock ``On the limited memory bfgs method for large scale optimization,''
\newblock {\em Mathematical programming}, vol. 45, no. 1, pp. 503--528, 1989.

\end{thebibliography}

\end{document}